\documentclass[fleqn,usenatbib]{mnras}
\usepackage{newtxtext,newtxmath}
\usepackage[T1]{fontenc}
\usepackage{ae,aecompl}
\usepackage{graphicx}	
\usepackage{amsmath}	
\usepackage{amssymb}	
\usepackage{natbib}
\usepackage{textcomp}
\usepackage{hyperref}
\usepackage{graphics}
\usepackage{longtable}[1]
\usepackage{lscape}
\usepackage{color}
\usepackage{xcolor}
\usepackage{rotating}
\usepackage{journal_names}
\usepackage{multicol,float}
\title[A remnant radio AGN in the EN1 field]{J1615+5452: a remnant radio galaxy in the ELAIS-N1 field}

\author[Z. Randriamanakoto et al.]{Z. Randriamanakoto$^{1}$\thanks{E-mail: zara@saao.ac.za},
C.\,H. Ishwara-Chandra$^{2}$, A.\,R. Taylor$^{3,\,4}$ \\
$^{1}$South African Astronomical Observatory, P.O. Box 9, Observatory 7935, South Africa\\
$^{2}$National Centre for Radio Astrophysics, TIFR, Post Bag No. 3, Ganeshkhind Post, 411007 Pune, India\\
$^{3}$Inter-University Institute for Data Intensive Astronomy, and Department of Astronomy, University of Cape Town,\\
      Private Bag X3, Rondebosch 7701, South Africa\\
$^{4}$Inter-University Institute for Data Intensive Astronomy, and Department of Physics and Astronomy, \\
      University of the Western Cape, Private Bag X17, Bellville 7535, South Africa     
}

\date{Accepted 2020 June 16. Received 2020 June 16; in original form 2019 November 30}
\pubyear{2020}
\hypersetup{draft}
\begin{document}
\label{firstpage}
\pagerange{\pageref{firstpage}--\pageref{lastpage}}
\maketitle

\begin{abstract}
We report the discovery of a remnant radio AGN J1615+5452 
in the field of ELAIS-N1. GMRT 
continuum observations at 150, 325 and 610\,MHz combined with archival data from the 1.4\,GHz NVSS survey
were used to derive the radio spectrum of the source. At a redshift $z \sim$ 0.33, J1615+5452 has a 
linear size of $\sim$\,100\,kpc
and spectral indices ranging between $\alpha^{1400}_{610} < -1.5$ and $\alpha^{325}_{150} = -0.61 \pm 0.12$.
While the source has a diffuse radio emission at low frequencies, 
we do not find evidence of core, jets or hotspots in the 1.4\,GHz VLA data of $\sim 5$\,arcsec angular resolution.
Such morphological properties coupled with a 
curved radio spectrum 
suggest that the AGN fueling mechanisms undergo a shortage of energy supply which is typical of a dying radio AGN. This is 
consistent with the observed steep curvature in the spectrum $\Delta\alpha \approx -1$, the estimated synchrotron age of $t_{\rm s}=76.0\,^{+7.4}_{-8.7}$\,Myr and 
 a $t_{\rm off}/t_{\rm s}$ ratio of $\sim 0.3$.

\end{abstract}


\section{Introduction}

In the active stage which usually lasts $\sim$\,$10 - 100$\,Myr \citep{1986MNRAS.219..575C}, 
the classical morphology of a radio galaxy is characterized by 
the presence of a core, a pair of lobes and including jets, and/or hotspots. 
Such features indicate the continuous injection (CI) of relativistic electrons that fuel
the active galactic nuclei (AGN). Following the active phase of the radio AGN,
the source enters the so-called remnant or dying phase, as
the nuclear engine switches off and the compact radio components, 
typical signatures of current activity, eventually disappear \citep{1999A&A...344....7P,2001AJ....122.1172S,2010ASPC..427..337K,2017NatAs...1..596M}. 

Despite the challenges of detecting radio AGNs in their remnant phase, especially prior to the era of low frequency radio facilities such as the Giant Metrewave Radio Telescope (GMRT, \citealp{1991ASPC...19..376S}), the Low-Frequency ARray (LOFAR, \citealp{2013A&A...556A...2V}) and the Murchison Widefield Array (MWA, \citealp{2009IEEEP..97.1497L,2013PASA...30....7T}),
\citet{1987MNRAS.227..695C} made a breakthrough when reporting the first discovery of a dying AGN 
known as B2 0924+30. Since then, this prototype of genuine radio remnants hosted by IC\,2476
has been the subject of follow-up studies 
\citep{2004A&A...427...79J,2017A&A...600A..65S,2018MNRAS.476.2522T}.
Meanwhile, other sources of this type have now been found in wide-field radio surveys 
\citep[e.g.][]{2007A&A...470..875P,2011A&A...526A.148M,2012ApJS..199...27S,2014MNRAS.440.1542D,2016A&A...585A..29B,2017A&A...606A..98B,2018MNRAS.475.4557M,2019PASA...36...16D}. 

The switching off of the central AGN activity is translated into a steep spectrum
$\alpha < -1.3$ (S\,$\propto$\,$\nu^{\alpha}$, e.g. \citealt{1994A&A...285...27K,2007A&A...470..875P,2017A&A...600A..65S})
of the associated radio emission in the GHz frequency regime, according to radiative ageing models \citep[e.g.][]{1962SvA.....6..317K,1970ranp.book.....P,1973A&A....26..423J}.  
This is because the injected relativistic electrons lose energy with time due to both synchrotron emission and Inverse Compton scattering with the Cosmic Microwave Background (CMB) photons \citep{1994A&A...285...27K}.   
Since higher energy electrons lose energy more quickly and have shorter radiative lifetimes, 
this results in a high-frequency turn down of the synchrotron spectrum occuring beyond a break 
frequency $\nu_{\rm b}$ that drifts in time to lower frequencies.
Below the break frequency, the radio spectrum of the non-active source has a spectral injection 
index, $\alpha_{\rm inj}$, typically in the range $< -0.5$ to $-1$ \citep[e.g.][]{1978ApJ...221L..29B,2011A&A...526A.148M,2016A&A...585A..29B}, 
while above $\nu_{\rm b}$, the spectral index $\alpha$ 
is steeper than $\alpha_{\rm inj}\,-\,0.5$
\citep{1962SvA.....6..317K, 1970ranp.book.....P, 1973A&A....26..423J}. 

Searches for remnant radio AGNs are generally based on detection of 
ultra-steep spectra at low frequencies
\citep[e.g.][]{2007A&A...470..875P,2007AJ....134.1245C}. 
To recover candidates with a spectrum not steep enough at low frequencies but 
ultra-steep above $\sim$\,1.4\,GHz, \citet{2011A&A...526A.148M} and \citet{2016A&A...585A..29B,2017A&A...606A..98B}, 
following the idea of \citet{2003A&A...404..133S}, 
used the radio spectral curvature parameter (SCP). 
This parameter uses multi-frequency information to
examine the difference between high frequency and low frequency spectral indices,
i.e.\, $\Delta \alpha = \alpha_{\rm high} - \alpha_{\rm low}$, and
its value correlates with the different evolutionary stages of the radio galaxy. 
In the typical case of a remnant AGN, $\Delta \alpha \ll -0.5$ \citep{2011A&A...526A.148M}.

Since remnant AGNs represent the final stage in the evolution of a radio galaxy, they are 
important objects for understanding of the radio galaxy life cycle, i.e.\ from the triggering of relativistic jets to the nuclear engine switch off, possibly followed by a restarting activity of AGN that may co-exist with a surrounding fossil radio emission (\citealp{2009BASI...37...63S}; \citealp{2011A&A...526A.148M}; \citealp{2012ApJS..199...27S}; \citealp{2012A&A...545A..91S}; \citealp{2013MNRAS.430.2137K}; \citealp{2017NatAs...1..596M}; \citealp{2018MNRAS.475.4557M}; \citealp{2020arXiv200409118J}). 
Characterizing these elusive objects 
through their morphology and spectra
will help determine the duty cycle of the radio core activity and the dynamical evolution of radio galaxies since mechanisms such as adiabatic expansion and radiative losses are expected to regulate the evolution of the remnant radio plasma once the radio jets switch off.
Thorough investigations of the AGN duty cycle and its dynamics are thus crucial to put constraints on the radio galaxy evolution models \citep[e.g.][]{2017A&A...606A..98B,2018MNRAS.475.4557M,2020MNRAS.tmp.1303S}. An extensive analysis of these sources will also help to address one of the key questions in the field of galaxy evolution: the co-evolution process between the supermassive black hole (SMBH) and its host galaxy  \citep[e.g.][]{1998A&A...331L...1S,2013ARA&A..51..511K,2015MNRAS.452..575S}. Although the tight correlation between the SMBH growth and the galaxy build-up has already been widely acknowledged, the impacts of the AGN physical properties and AGN feedback on the cosmic evolution of the host galaxy and its environment are still under debate (\citealp{2011ApJ...734...92J}; \citealp{2012ARA&A..50..455F} and references therein). 

Because of the rapid timescale of particle energy decay, and the paucity of high sensitivity 
data at multiple wavelengths, only a handful of steep spectrum dying radio sources have been observed, especially in the cm wavelength regime. 
However, in addition to their spectral signature, fossil AGNs may be recognizable
in deep low-frequency images as relatively bright and diffuse emission devoid of a core and/or hotspots and with resolutions of a few arcseconds. Such emission arises from low energy particles that are relatively unaffected by the fast spectral evolution \citep{2007A&A...470..875P,2011A&A...526A.148M,2017A&A...606A..98B,2017NatAs...1..596M}. 

In this work, we report the discovery of a new remnant radio galaxy, J1615+5452, in the field of the European Large-Area ISO Survey-North\,1 \citep[ELAIS-N1 or EN1,][]{2000MNRAS.316..749O} from the GMRT 610\,MHz observations. We use the archival low-frequency GMRT observations at 150\,MHz and 325\,MHz and 1400\,MHz data from the Karl G. Jansky Very Large Array (VLA) to obtain the source radio spectrum and to reconstruct its history of AGN activity. 

The paper is structured as follows. We describe the observations and data processing in Section\,\ref{sec:data}. Radio morphology and integrated spectra of the source are reported in Section\,\ref{sec:radio-prop} and Section\,\ref{sec:discuss} investigates the nature of the remnant AGN. We discuss the results in Section\,\ref{sec:discuss2} and provide a summary of the work in Section\,\ref{sec:conclusion}. We adopt the following cosmological parameters throughout the paper: $H_0 = 67.8\,{\rm km~s^{-1}~ Mpc^{-1}}$, $\Omega_{\rm m} = 0.308$, and $\Omega_{\rm \Lambda} = 1 - \Omega_{\rm m}$ \citep{2016A&A...594A..13P}.

\section{Observations and data processing}\label{sec:data}

EN1 is a field in the northern part of the sky centred near ${{\rm RA} = \rm 16^{h}10^{m},\,{\rm DEC} =
+54^{\circ}35'}$\,(J2000 coordinates). The low foreground infrared emission in this region has made it a primary target for deep extragalactic surveys, including deep radio continuum surveys over a range of frequencies \citep{2008MNRAS.383...75G,2009MNRAS.395..269S,2010ApJ...714.1689G,2011ApJ...733...69B,2014A&A...568A.101J,2016MNRAS.459L..36T,2019MNRAS.490..243C,2020MNRAS.491.1127O}.  This work uses observations of EN1 with the GMRT taken at 150, 325, and 610\,MHz as well as data at 1400\,MHz from the VLA. Table\,\ref{tab:log-obs} summarizes the GMRT observation logs of the field including the central frequency, the bandwidth, the observing date and the on-source integration time.

\begin{table*}

\caption{\small Log of the GMRT observations of the ELAIS-N1 field.}

\centering
\scalebox{0.85}{
  \begin{tabular}{cccccccc}
  \hline
  \hline
Band & Central Fqcy  &  BW     & Project Code \& PI &  Obs.date & On-source time & Resolution & $\sigma_{\rm rms}$   \\
${\rm [MHz]}$ & ${\rm [MHz]}$ &  [MHz]  &                    &           & [hr]     & [arcsec]   & [$\mu$Jy/beam]  \\
      (1)     &  (2)    & (3)                &   (4)     &   (5)    & (6)        & (7) & (8)    \\ \hline

150 & 153 &     6      & 12MDB01, Dennefeld      &   May 2007   & 17 x 0.5 & $38\times21$  & 3500  \\
325 & 323        & 32      & ~~~24\_026, Wadadekar   &   May  2013  & 2  x 8.5 &  $9$  &  70 \\
610 & 610        & 32      & ~~~19\_064, 20\_020, 21\_086, 22\_057, Taylor      &  2011 - 2017 & 3  x 1   &  $6$  &  40\\
\hline
\multicolumn{8}{@{} p{18.5cm} @{}}{\footnotesize{\textbf{Notes. }
Columns 1 \& 2: nominal and precise central frequencies; 
Column 3: bandwidth used for mapping the field;
Column 4: GMRT project code and name of the principal investigator;
Column 5: date of the observations;
Column 6: on-source integration time;
Columns 7 \& 8: angular resolution and rms noise of the radio image.
}}
\end{tabular}
}
\label{tab:log-obs}
\end{table*}

\subsection{GMRT observations}

The 610\,MHz data is from the ELAIS-N1 wide area survey of this region using the GMRT. This is a large area survey of 12.8 sq.\,degrees comprised of 51 pointings of 3 hours each and observed between 2011 and 2017 (Ishwara-Chandra et al.\ 2020, submitted). We used the GMRT Software Backend (GSB) with a 32\,MHz bandwidth split into 256 spectral
channels to avoid bandwidth smearing. On most days, we observed 3C286 as a primary calibrator while 3C48 for flux and bandpass calibration observed over a few days only. The data analysis was carried out using a fully automated CASA-based pipeline which is described in detail in Ishwara-Chandra et al.\ (2020, submitted). In a nutshell, the data was flagged, gain calibrated, and after which several rounds of self-calibration on target along with flagging on residuals were being processed. For imaging, we used a robust parameter of 0 in the Briggs weighting scheme. The final rms noise was $\sim40\,\mu$Jy/beam at a resolution of 6\,arcsec circular beam. J1615+5452 was discovered from this deep image as fuzzy source without any hotspots or compact feature. The angular extent of this source is much smaller than the largest source that can be detected at this frequency (17$'$).

The GMRT 325\,MHz observations cover 3.6 square degree area of the ELAIS-N1 field with a phasecenter  set at ${\alpha = \rm 16^{h}10^{m},\,\delta = +54^{\circ}40'}$. The survey is part of the GMRT radio continuum imaging of the Herschel/HerMES field (PI:\,Wadadekar, Prop\,ID:\,24\_026). Two single pointing observations of $\approx$\,8.5\,hours each were carried on 2013 May\,26 \& 27 with a total bandwidth of $\sim$\,32\,MHz. Each $\sim$\,52\,min scan of the target field was followed by a $\sim$\,5min secondary calibrator observation to track the
ionospheric conditions, the RFI contamination and any variations in the receiver system throughout the entire observing run. We have processed this archival data through the same CASA-based pipeline used to analyse the wide-area 610\,MHz data. 3C286 and 3C48 were used for flux calibration and 1459+716 for phase calibration.  Each day's data was calibrated separately and the split files of target field were combined before imaging with CASA {\tt tclean} using the default robust weighting of 0. We  performed four rounds of phase-only self-calibration followed by five rounds of amplitude and self-calibration, which included residual-based flagging in between each round. The final image reached a sensitivity down to $\sigma \approx 70\,\mu$Jy/beam without applying direction-dependent calibration. Primary beam correction was done following standard procedures. J1615+5452 is clearly detected in this image and the source is much smaller than the largest angular size that can be imaged using the GMRT at 325\,MHz ($\sim$ 30$'$).

We searched for J1615+5452 in the 150\,MHz TIFR GMRT Sky Survey \citep[TGSS,][]{2017A&A...598A..78I} and the source was not clearly detected.
The nearest survey pointing center is 1.6\,degree away which is close to the half-power beamwidth at 150\,MHz. There is a faint source at the expected location with a peak flux density of about 20\,mJy and a total flux density of 95 $\pm$ 21\,mJy. Since the peak flux is less than 5 times the local rms, it did not get listed in the TGSS catalogue. 
We searched for other 150\,MHz archival data with the GMRT and found  long observations with $\sim$ 8.5\,hours on-source time, taken in May\,2007 (PI: Dennefeld). The signal bandwidth was 6\,MHz with a correlator bandwidth of 8\,MHz split into 128 channels to minimise the effects of bandwidth smearing. 3C147 and 3C48 were observed for primary calibration and 1459+716 for secondary calibration. The data was analysed using the {\tt SPAM} pipeline \citep{2009A&A...501.1185I} following standard procedure which included direction dependent calibration for total intensity continuum imaging (robust = 0).
The rms noise of the image is $\sim$ 3.5 mJy/beam with a resolution of 37.7 $\times$ 20.9 arcseconds at the position angle of 52\,deg. The source is clearly detected in this image which is the one we use for the flux density estimate at 150\,MHz (Table\,\ref{tab:prop-sources}).

\subsection{VLA observations at 1.4 GHz}

A snapshot image of J1615+5452 is retrieved from the NRAO VLA Sky Survey \citep[NVSS,][]{1998AJ....115.1693C}. At a resolution of $\sim\,45$\,arcsec coupled with a sensitivity of $\sim$\,2.5\,mJy/beam above 5$\sigma$, the 1.4\,GHz survey records a flux density of 5.6\,mJy for this given source. We also retrieved the radio image from the Faint Images of the Radio Sky at Twenty-Centimeters survey \citep[FIRST,][]{1995ApJ...450..559B},
but the NVSS point-like source is only detected in FIRST below 4$\sigma$ with a very weak intensity $<$\,0.4\,mJy. 
We did not find radio counterparts as well in the 10 sq. degrees 1.4\,GHz VLA image of the EN1 field 
published by \citet{2011ApJ...733...69B}. The local rms noise is $\sim90\,\mu$Jy/beam. Unlike NVSS, these two VLA observations are at a higher angular resolution of $\sim5$\,arcsec. No compact feature at high resolution suggests the absence of core emission.

\section{Properties of J1615+5452}\label{sec:radio-prop}

\subsection{Host galaxy}\label{sec:host}

\begin{figure}
{\resizebox{1.00\hsize}{!}{\includegraphics[trim= 0.cm 0.cm 0.cm 0.cm, clip]{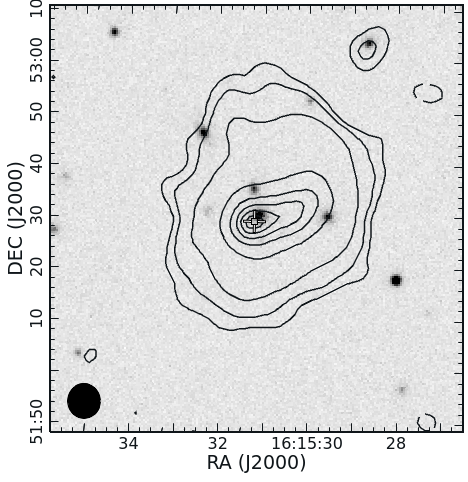}}}
\caption{\small The 610\,MHz radio continuum contours (levels $-$2, 2, 3, 5, 10, 30, 40, 45, 48, 50 $\times$ $\sigma$\,(40 $\mu$Jy/beam)) overlaid on the SDSS-DR12 $i$-band image. The cross is the centroid of the radio emission. The filled ellipse in the lower left represents the FWHM contour of the synthesized beam at 610\,MHz.}
\label{fig:optical_overlay}
\end{figure}

Figure\,\ref{fig:optical_overlay} shows the 610\,MHz radio continuum contours of J1615+5452 overlaid on a $i$-band image from the Sloan Digital Sky Survey Data Release 12
\citep[SDSS-DR12,][]{2015ApJS..219...12A}. A red spheroidal galaxy is located coincident with the peak of the radio emission and slightly $\sim1.9$\,arcsec East of the centroid of the radio source. 
Based on spectroscopic observations from the SDSS-DR14 \citep{2018ApJS..235...42A}, this galaxy has a redshift of $z = 0.32936 \pm 0.00005$ at which 1\,arcsec corresponds to 4.89\,kpc.
With no obvious nebular emission lines from young blue stars, its integrated spectrum 
retrieved from the SDSS Catalog Archive Server\footnote{\href{https://skyserver.sdss.org/dr12/en/tools/chart/navi.aspx}{https://skyserver.sdss.org/dr12/en/tools/chart/navi.aspx}} is rather characterized by a prominent break around 4000\,\AA~and the presence of Ca H and K lines along with other strong absorption lines like MgI and NaD. Such features are typical of an early-type elliptical galaxy \citep{1992ApJS...79..255K}.

The newly detected remnant radio emission is likely associated with the SDSS galaxy. In fact, the slight misalignment between the position of the optical galaxy and the peak of the radio emission is consistent with a host galaxy that is moving away from the relic radio plasma after its AGN activity ceases
\citep[e.g.][]{,2019PASA...36...16D}. Such an offset has also been reported by \citet{2007ApJ...659..225G} when studying 3C338, a restarted radio galaxy where the remnant plasma of the source is clearly disconnected from the host galaxy. 

The potential host galaxy does not belong to any cluster of galaxies detected in EN1 field (V. Parekh, private communication). However, there are three SDSS galaxies with $g<22$\,mag within its 0.5$'$ radius. Such a proximity to other galaxies suggests that the host galaxy could reside in a loose group environment.
It is worth investigating the nature of the galaxy environment since \citet{2011A&A...526A.148M} reported that fossil lobes of cluster-based remnant radio galaxies tend be more confined than the ones associated with a dying radio source found in the field. Unfortunately, no X-ray  observations covered this area of the sky to help investigate the effects of the surrounding intercluster medium on the radio source.

\begin{table}
  \centering
\caption{\small Properties of the peculiar radio source.}
  \begin{tabular}{ll}
  \hline
  \hline
Source name          & J1615$+$5452 \\
RA (J2000)           & 16h15m31.1s  \\
DEC (J2000)          & $+$54d52m28s   \\ 
{\it z}              & 0.32936\\\\

$S_{150}$ 	         & 81.1 $\pm$ 6.3\,mJy \\
$S_{325}$            & 50.5 $\pm$ 2.5\,mJy \\
$S_{610}$            & 24.8 $\pm$ 1.2\,mJy  \\
$S_{1400}$           & ~6.7 $\pm$ 0.8\,mJy\\\\

$\alpha^{1400}_{150}$& $-$1.12$\pm$ 0.06 \\
$\alpha^{1400}_{610}$& $-$1.58$\pm$ 0.15 \\
$\alpha^{610}_{325}$ & $-$1.12$\pm$ 0.11\\
$\alpha^{325}_{150}$ & $-$0.61$\pm$ 0.12\\
$\Delta\alpha$ (SCP) & $-$0.97$\pm$ 0.19\\\\

$L_{1.4\,{\rm GHz}}$  & $2.64 \times 10^{24}\, {\rm W\,Hz^{-1}}$ \\

\hline 
\end{tabular}
\label{tab:prop-sources}
\end{table}

\begin{figure*}
\centering
\begin{tabular}{ccc}
{\resizebox{0.44\hsize}{!}{\includegraphics{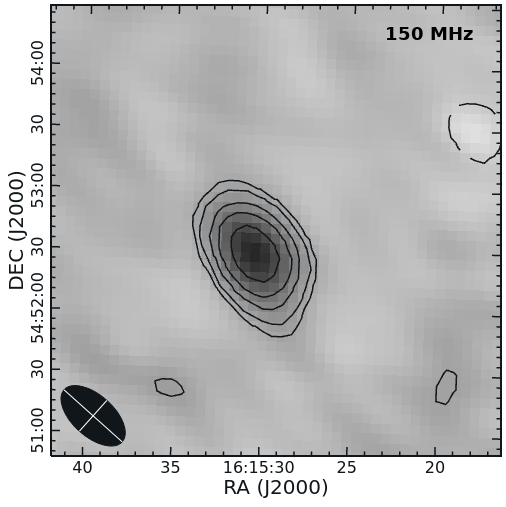}}}
{\resizebox{0.44\hsize}{!}{\includegraphics{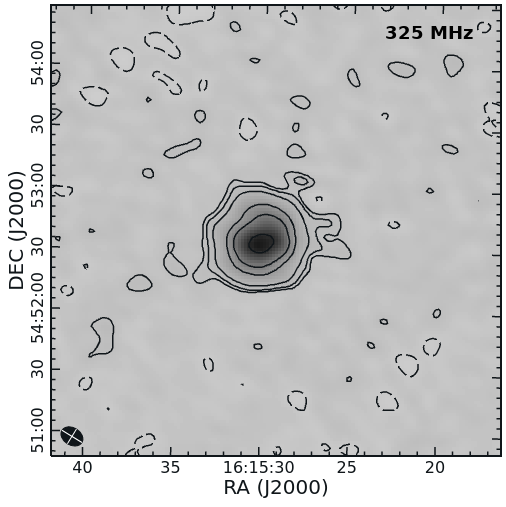}}}\\
{\resizebox{0.44\hsize}{!}{\includegraphics{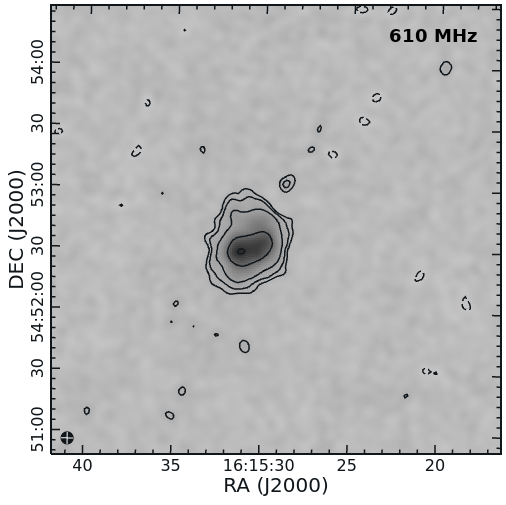}}}
{\resizebox{0.44\hsize}{!}{\includegraphics{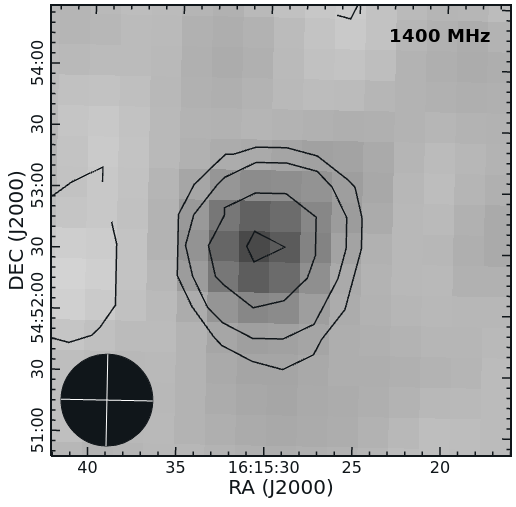}}}\\

\end{tabular}
\caption{\small Radio maps of the source J1615+5452 with the synthesized beam shown in the bottom left corner. Overlaid are the total intensity contours: ({\em top left}) GMRT 150\,MHz map, levels $-$2, 2, 3, 5, 7, 10 $\times$ $\sigma$\,(3.1 mJy/beam); ({\em top right}) GMRT 325\,MHz map, levels $-$3, 3, 5, 10, 30, 50, 100 $\times$ $\sigma$\,(70 $\mu$Jy/beam); ({\em bottom left}) GMRT 610\,MHz map, levels $-$3, 3, 5, 10, 30, 50 $\times$ $\sigma$\,(40 $\mu$Jy/beam); ({\em bottom right}) NVSS 1400\,MHz map, levels $-$2, 2, 3, 5, 7, $\times$ $\sigma$\,(0.34 mJy/beam).}

\label{fig:radio-images}
\end{figure*}

\subsection{Morphology and flux densities}\label{sec:fluxes}
Figure\,\ref{fig:radio-images} shows the radio images at all four frequencies. As noted in Table\,\ref{tab:log-obs}, the GMRT angular resolution ranges from 6\,arcsec at 610\,MHz to 20\,arcsec at 150\,MHz. We used {\tt PyBDSF} \citep{2015ascl.soft02007M} to measure the flux density of the source from all the available images. The derived values are presented in Table\,\ref{tab:prop-sources} with the associated errors $\sigma_{\rm I}$ computed following the expression outlined by \citet{2003AJ....125..465H} but with a slight modification to account for our instrumental and pointing errors:
\begin{equation}\label{eq:dS}
{\rm \sigma_I} = I\,\sqrt{ 2.5\left(\frac{\sigma}{I} \right)^{\!\!2} +  0.05^2}
\end{equation}
where $I$ is the integrated flux density and $\sigma$ the local rms noise.

We first noticed the peculiarity of J1615+5452 by cross-matching the GMRT 610\,MHz radio image of EN1 with
the high resolution 1.4\,GHz VLA data. While the source is easily detectable at 610\,MHz with a projected linear size of $\sim$\,92\,kpc (0.31\,arcmin) at $z \sim\,0.33$ and 
a flux density of $S_{610} = 24.8 \pm$ 1.2\,mJy, no 1.4 GHz contours however,
have been recovered. The VLA high resolution images did not reveal any compact core, jets or hotspots. 
Such a non-standard morphological structure is worth an investigation as it  was also observed by e.g. \citet{2016A&A...585A..29B} when studying a remnant radio galaxy 
in a particularly low-density environment. 

The radio emission is also featured at 150\,MHz with a flux density of $S_{150} = 81.1 \pm$ 6.3\,mJy. In the 325\,MHz image, 
the source has extended diffuse emission with $S_{325} = 50.5~\pm$ 2.5\,mJy and a projected linear size of 100\,kpc (0.34 arcmin). The source extraction with {\tt PyBDSF} in the low resolution NVSS data resulted in a flux density of $S_{1400} = 6.7~\pm$ 0.8\,mJy which translates to a radio luminosity of 
$L_{1.4\,{\rm GHz}} = 2.64 \times 10^{24}\, {\rm W\,Hz^{-1}}$.

We also estimated an upper limit of the radio core prominence (CP) which is a parameter used by e.g. \citet{1988A&A...199...73G}, \citet{2016MNRAS.462.1910H}, \citet{2016A&A...585A..29B, 2017A&A...606A..98B}, \citet{2018MNRAS.475.4557M} in their search for candidate remnant AGNs. This was done by taking the ratio between the $3\sigma$ local rms noise in the high resolution VLA image from \citet{2011ApJ...733...69B} and the total flux density at 150\,MHz. We derived a value of CP $\lesssim 3.3 \times 10^{-3}$. However, given that this value changes depending on the observing frequencies and that a low core prominence is expected for a flat-spectrum core component combined with a steep-spectrum of the total radio emission \citep{2008MNRAS.388..176H}, we also considered an alternative method to investigate the AGN activity in the core. By taking the value of $L_{1.4\,{\rm GHz}}$, we used the following empirical relation established by \citet{1990A&A...227..351D}:
\begin{equation}\label{eq:CPval}
{\rm log\,CP} = -0.55\,{\rm log}\,L_{1.4\,{\rm GHz}} + 11.76     
\end{equation}
to predict a much higher value of CP\,$\sim$\,0.02. The \citet{1990A&A...227..351D} relation states that the CP is inversely proportional to the radio power, i.e. low luminosity radio sources (with radio powers smaller than $10^{26}\,{\rm W\,Hz^{-1}}$) are expected to have a relatively higher CP compared to the high luminosity ones. 

The values of the core prominence and the absence of the compact radio structure suggest that J1615+5452 could be a dying AGN (see Section\,\ref{sec:discuss2}). The analysis of the source radio spectrum in 
Section\,\ref{sec:radio-spectra} will help determine whether the AGN activity has indeed switched off.

\subsection{Radio spectrum}\label{sec:radio-spectra}
In the active stage, a broken power-law fit also known as the CI model \citep{1962SvA.....6..317K,1973A&A....26..423J} best describes the radio source spectrum with a spectral index $\alpha \sim \alpha_{\rm inj}$ below the break frequency and $\alpha < \alpha_{\rm inj} - 0.5$ above $\nu_{\rm b}$. If the source is young enough and the observing frequency range is limited to $\sim$\,GHz frequencies, then a simple power-law with $\alpha \sim \alpha_{\rm inj}$ produces a reasonable fit to its radio spectrum still unaffected by the radiative losses.

\begin{figure}
\centering
\begin{tabular}{ccc}
{\resizebox{0.95\hsize}{!}{\includegraphics[trim= 0cm 0cm 0cm 0cm, clip]{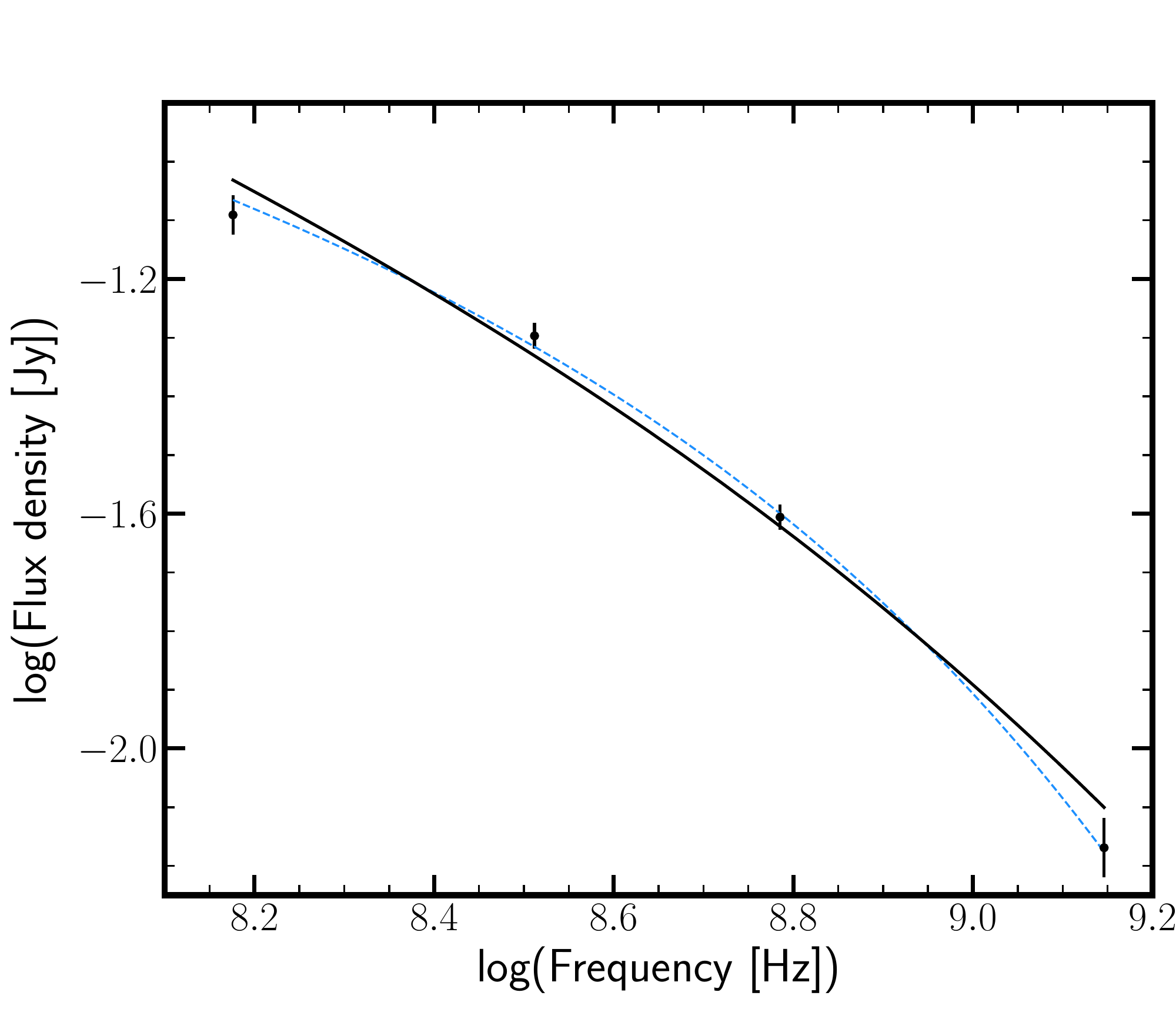}}}
\end{tabular}
\caption{The integrated radio spectrum of the candidate remnant radio AGN J1615+5452. The dashed line represents the best fit CI$_{\rm off}$ model (though associated with a relatively flat injection index of 0.40). For $\alpha_{\rm inj} = 0.61$ and $B_{\rm eq} = 4.1\, \mu$Jy, the same model returns the solid line with an observed break frequency $\nu_{\rm b} \sim 384$\,MHz and an off component break at 5\,GHz.}
\label{fig:spectrum-S1}
\end{figure}

For J1615+5452, the best fit model\footnote{The best fit CI$_{\rm off}$ model by \citet{1994A&A...285...27K} is briefly described in Section\,\ref{sec:age}.} of the integrated spectrum (see Figure\,\ref{fig:spectrum-S1}) is a broken power-law fit combined with an exponential break frequency $\nu_{\rm b,\,exp}$. At low frequencies, the spectral shape is typical of active radio galaxies with an index of  $\alpha^{325}_{150} = -0.61\pm$ 0.12. Steep spectra with $\alpha^{610}_{325} = -1.12\pm$ 0.11 and $\alpha^{1400}_{610} = -1.58\pm$ 0.15 are, however, recorded above a certain spectral break likely occurring between 325 and 610\,MHz. The value of such a parameter and the resulting synchrotron age are derived in Section\,\ref{sec:age}. 

In the frequency range 150 - 1400\,MHz,
$\alpha^{1400}_{150} = -1.12\pm$ 0.06 and the spectral curvature parameter $\Delta\alpha = \alpha^{1400}_{610} -
\alpha^{325}_{150}$ is equal to $-0.97 \pm 0.19$. The measured values of the SCP and the spectral indices are reported in Table\,\ref{tab:prop-sources}.

The spectral break at low frequencies, the steep high frequency spectra as well as a SCP $\ll-0.5$ imply the predominance of nonthermal synchrotron emission and the central engine switch off. Such characteristic features of the spectra classify once again J1615+5452 as a dying radio AGN \citep[e.g.][]{2007A&A...470..875P,2011A&A...526A.148M,2016A&A...585A..29B}. 

\section{Investigating the dying radio AGN}\label{sec:discuss}

\subsection{Source energetics}
A first order approximation of the synchrotron age requires an estimate of the magnetic field strength. We assume minimum energy conditions between particles and magnetic field to derive the equipartition magnetic field $B_{\rm eq}$ which is given by 
\begin{equation}
B_{\rm eq}[{\rm G}] = \Big(\frac{24\pi}{7}\,u_{\rm min}\Big)^{1/2}.
\end{equation}

Assuming an uniform magnetic field and
an isotropic particle distribution, we computed the minimum energy density $u_{\rm min}$ following the \citet{2004IJMPD..13.1549G} formula:
\begin{equation}\label{eq:umin}
\begin{split}
u_{\rm min}\Big[{\rm \frac{erg}{cm^3}}\Big] = \xi(\alpha, \nu_1, \nu_2)(1 + k)^{4/7}(\nu_{0}\,[{\rm MHz}])^{-4\alpha/7} \\
          \times~(1 + z)^{(12-4\alpha)/7}\Big(I_{0}\,[{\rm \frac{mJy}{arcsec^2}}]\Big)^{4/7}(d\,{[{\rm kpc}]})^{-4/7}
\end{split}
\end{equation}
where the parameter $\xi$ is dependent on frequencies ($\nu_1$ = 10\,MHz and $\nu_2$ = 100\,GHz) and 
the synchrotron spectral index $\alpha < 0$  presumed for the observed spectrum;
$k$ is the relativistic proton-to-electron energy density ratio.
$I_0$ is the source surface brightness measured at the frequency $\nu_0$ and $d$ is the source depth.

At $\alpha = -0.61$ with $\nu_0$ = 325\,MHz, we find $u_{\rm min} = {\rm 1.54 \times 10^{-12}\,erg\,cm^{-3}}$
and $B_{\rm eq} \sim 4.1\,\mu$G. The other parameters in Eq.\,\ref{eq:umin} are as follows: $k = 1$, $I_0$ = 0.43\,mJy/arcsec$^2$, and $d \sim 100$\,kpc (the average of the angular extents of the source). Note that the value of $\alpha$ is consistent with the observed injection index $\alpha_{\rm inj}$ measured at $\nu_0$ where the steepening is still not affecting the shape of the radio spectrum. Regarding the frequency interval, the choice of a relatively low value of $\nu_1$ = 10\,MHz is motivated by the importance of the energy density of relativistic electrons radiating their energy at lower frequencies \citep[e.g.][]{1970ranp.book.....P, 2005AN....326..414B}. The value of $\nu_2$ is however less critical for the minimum energy estimates. In fact, if $\nu_2$ = 10\,GHz for $\alpha = -0.61$, $u_{\rm min}$ only decreases by 9\,percent from its current value.

Previous works have reported that the classical formula used to calculate $B_{\rm eq}$ may underestimate the magnetic field strength because of inadequate assumptions of the frequency integration limits \citep[e.g.][]{1997A&A...325..898B,2005AN....326..414B}. A revised estimate is:  
\begin{equation}
B^{'}_{\rm eq}\,[{\rm G}] \sim 1.1 \gamma_{\rm min}^{\frac{1+2\alpha}{3 - \alpha}} B_{\rm eq}^{\frac{7}{2(3-\alpha)}}
\end{equation}
where $B^{'}_{\rm eq}$ is a modified equipartition magnetic field based on a limit of the Lorentz factors $\gamma$ assuming $\gamma_{\rm max} \gg \gamma_{\rm min}$. For $\gamma_{\rm min}$ = 100 and $\gamma_{\rm max}$ = 10$^6$, we find $B^{'}_{\rm eq} \sim 4.9\,\mu$G using the 325\,MHz radio emission. Because of the flat radio spectrum at low  frequencies, we only expect $\sim$\,$20 - 30$\,percent deviation of the revised value from the classical $B_{\rm eq}$ \citep{2005AN....326..414B}. In this work, such a variation is only around 16\,percent. We thus adopt the magnetic field strength $B_{\rm eq} \sim 4.1\,\mu$G in the remainder of this work.

\subsection{Synchrotron age}\label{sec:age}

With estimates of the magnetic field strength $B_{\rm eq}$ and the break frequency $\nu_{\rm b}$, one can calculate to first order the spectral age $t_{\rm s}$ using the following expression \citep{2011A&A...526A.148M}:
\begin{equation}\label{eq:age}
t_{\rm s}\,[{\rm Myr}] = 1590 \bigg[\frac{(B_{\rm eq}\,[\mu {\rm G}])^{1/2}}{(B_{\rm eq}^2 + (B_{\rm IC}\,[\mu {\rm G}])^2)(1 + z)^{1/2}\,(\nu_{\rm b}\,{[{\rm GHz}]})^{1/2}}\bigg]
\end{equation}

where $B_{\rm IC} = 3.25\,(1+z)^2$ is the Inverse Compton equivalent magnetic field. By setting a lower limit to the break frequency such that $\nu_{\rm b} = 610$\,MHz, an upper limit to the source age is in the order of 120\,Myr. 

We also run {\tt BRATS}\footnote{The Broadband Radio Astronomy ToolS is a software package that uses spectral ageing models to derive the 
properties and morphology of a radio source. For more details,
visit \url{http://www.askanastronomer.co.uk/brats.}} \citep{2013MNRAS.435.3353H,2015MNRAS.454.3403H}
to fit the radio spectrum and to ultimately get a more robust estimate of the synchrotron age. Given the remnant nature of the source,
we fit the KGJP model to the data. This remnant radio emission model developed by \citet{1994A&A...285...27K} is also known as the CI$_{\rm off}$ model and it is a modification of the standard continuous injection model \citep{1962SvA.....6..317K, 1973A&A....26..423J} to account for the AGN switch off. 

By considering an injection index $\alpha_{\rm inj} = -0.61$ (the spectral index at low frequencies), a minimum and maximum Lorentz factor of $\gamma_{\rm min}$ = 100 and $\gamma_{\rm max}$ = 10$^6$, a magnetic field strength $B_{\rm eq} = 4.1\,\mu$G, with the other input parameters set to their default values, {\tt BRATS} returns a fitted age of $t_{\rm s}=76.0\,^{+7.4}_{-8.7}$\,Myr, a break frequency $\nu_{\rm b} = 384$\,MHz and an off component break $\nu_{\rm b,\,exp} = 5029$\,MHz. The timescales during active ($t_{\rm on} = 54.9\,^{+7.4}_{-8.5}$\,Myr) and quiescent ($t_{\rm off} = \,21.0\,^{+0.0}_{-2.1}$\,Myr) 
phases were also computed. With $t_{\rm s} = t_{\rm on} + t_{\rm off}$, we get a $t_{\rm off}/t_{\rm s}$ ratio of $\sim 0.3$. The off component break frequency, $\nu_{\rm b,\,exp} = \nu_{\rm b}\,(t_{\rm s}/t_{\rm off})^2$, does not lie within the available frequency coverage because the source is observed soon after the radio core switch off according to the fitting model, i.e.\,the timescale of the quiescence phase is significantly shorter than that of the active phase \citep{2007A&A...470..875P}. In that case, $\Delta\alpha \sim -1$ (derived in Section\,\ref{sec:radio-spectra}) can only be considered as an upper limit of the spectral curvature parameter. 

The source age of 76\,Myr is in agreement with the source morphology and the steep spectral properties reported in Section\,\ref{sec:radio-prop}. Note, however, that the computed value is dependent on some critical parameters such as $\alpha_{\rm inj}$. For instance, the derived age falls between 37 - 106\,Myr while varying the injection index $\alpha_{\rm inj}$ between 0.5 and 0.8. Such a range is consistent with the upper limit to the source age as well as the typical ages (a few tens of Myr) for dying radio galaxies  \citep[e.g.][]{1978ApJ...221L..29B,2011A&A...526A.148M,2016A&A...585A..29B}. 

Although we can achieve the best fit model with $\alpha_{\rm inj} = 0.40$ and $B_{\rm eq} = 12.5\,\mu$G as free parameters ($\chi^2_{\rm red}$ = 1.42, dashed line in Figure\,\ref{fig:spectrum-S1}), we decided to consider {\tt BRATS} outputs associated with fixed values of $\alpha_{\rm inj} = 0.61$ and $B_{\rm eq} = 4.1\,\mu$G. The latter parameters are consistent with the radio properties of the source (e.g. the spectral index at lower frequencies) and the resulting model still gives a reasonable fit to the radio spectrum ($\chi^2_{\rm red}$ = 8.86, solid line in Figure\,\ref{fig:spectrum-S1}). Table\,\ref{tab:nub-chi} summarizes {\tt BRATS} fitting results in both cases.

\section{Discussion}\label{sec:discuss2}

A low core prominence may result from the very weak or no core emission, a potential evidence of the central AGN switch off. In the case of J1615$+$5452, the observed core prominence of $\lesssim 3.3 \times 10^{-3}$ falls within the range of CP $< 10^{-4}-~5 \times 10^{-3}$ that are associated with the remnant radio sources from the literature \citep[e.g.][]{1988A&A...199...73G,2016MNRAS.462.1910H, 2017A&A...606A..98B,2018MNRAS.475.4557M}. Such a low value is well below CP\,$\sim$\,0.02 (the CP value expected for a source of the same radio power assuming the empirical correlation found by \citealt{1990A&A...227..351D}) and it is in agreement with a quiescent nuclear engine or an active one but at a very weak level than what is observed in typical radio sources. Furthermore, the value of CP\,$\sim$\,0.02 also falls within the interval of 0.1 - 0.001, which is a CP range reported by \citet{1990A&A...227..351D} for B2 sources with radio powers between $10^{24} - 10^{26}\,{\rm W\,Hz^{-1}}$ including a very small fraction of remnant radio AGNs. 
Nevertheless, classifying a radio source as a remnant based only on its core activity is deemed unreliable, especially for lower resolution and sensitivity observations where a faint core may be missed \citep[e.g.][]{2018MNRAS.475.4557M}. It is thus recommended to combine this parameter with other selection criteria to achieve unbiased detection
\citep[e.g.][]{2017A&A...606A..98B}.

In this work,  
$\alpha^{1400}_{325} = -1.37 \pm 0.09$ and
$\Delta\alpha \sim -0.97 \pm 0.19$.
These spectral characteristics are comparable to those of the
complete samples of remnant radio galaxies drawn by \citet{2007A&A...470..875P} and \citet{2011A&A...526A.148M}. 
The former authors define a dying radio source based on steep spectral shape ($\alpha^{1400}_{325} < -1.3$) and
the absence of compact radio structures, whereas the latter  consider as well the source spectral curvature ($\Delta\alpha < -0.5$) as one
of its complementary selection criteria to search for candidate remnant AGNs. 

In addition, J1615+5452 has similar spectral properties with those of remnant
radio galaxies observed in the LOFAR Lockman Hole field ($\alpha^{1400}_{150} < -1.2$ and
$\Delta\alpha = \alpha^{1400}_{325} - \alpha^{325}_{150} < -0.5$, \citealt{2017A&A...606A..98B}). Note however that the source spectral index $\alpha^{325}_{125} = -0.61 \pm 0.12$ at low frequencies is consistent with the spectra of active radio galaxies. \citet{2011A&A...526A.148M} and \citet{2016A&A...585A..29B} also noticed such spectral shape while investigating the radio properties of B2 1619+29 and blob1 (a remnant radio galaxy in the LOFAR field), respectively. A slow rate of radiative cooling or a young plasma are suggested to explain the observed spectral feature. It is therefore crucial to carry multi-frequency analyses for an optimal detection of genuine dying radio sources with such spectra.

The computed magnetic field $B_{\rm eq} = 4.1\,\mu$G is in agreement with the magnetic field strengths of previously studied remnant radio AGNs \citep[e.g.][]{2011A&A...526A.148M,2014MNRAS.440.1542D,2016A&A...585A..29B,2019PASA...36...16D} and dying giant radio galaxies \citep[e.g.][]{2015MNRAS.453.2438T} of the order of a few $\mu$G. Such a weak magnitude implies the predominance of the Inverse Compton scattering from the CMB over the synchrotron losses throughout the radiative cooling of the diffuse plasma \citep{1994A&A...285...27K}.

The age range of the source in 
Section\,\ref{sec:age} is comparable to those of other dying AGNs such as B2 0924+30 (54\,Myr by \citealt{2004A&A...427...79J}, $\sim$\,88\,Myr by \citealt{2017A&A...600A..65S}, 78\,Myr by \citealt{2018MNRAS.476.2522T}), the median age of the remnant samples from the Westerbork Northern Sky Survey (63\,Myr, \citealt{2007A&A...470..875P}) and blob1 (75\,Myr, \citealt{2016A&A...585A..29B}). With $t_{\rm s} = 76$\,Myr, the source is relatively older than the low luminosity radio galaxies from the B2 sample with a median age of 31\,Myr \citep{1999A&A...344....7P}. 

\begin{table}
\caption{\small Fitting results from {\tt BRATS} ${\rm CI_{off}}$ model.}

\centering
\scalebox{0.9}{
  \begin{tabular}{ccccccc}
  \hline
  \hline
$\alpha_{\rm inj}$ & $B_{\rm eq}$ & $\nu_{\rm b}$  & $\nu_{\rm b,\,exp}$  & $t_{\rm s}$ & $t_{\rm off}/t_{\rm s}$ & $\chi^2_{\rm red}$ \\
 & [$\mu$G] & [MHz] & [MHz] & [Myr] & \\
(1) &  (2)    & (3)  &   (4)  & (5) & (6) & (7)  \\ \hline

0.40 & 12.5 & 347 & 1024 & $35.9\,^{+1.9}_{-3.5}$ & 0.6 & 1.42 \\
&&&&&\\
0.61 & 4.1 & 384 & 5029 & $76.0\,^{+7.4}_{-8.7}$ & 0.3 & 8.86 \\
\hline
\multicolumn{7}{@{} p{8cm} @{}}{\footnotesize{\textbf{Notes. }
Columns 1 \& 2: the injection index and the magnetic field as input parameters; Column 3: the break frequency; Column 4: the off-component break frequency; Columns 5 \& 6: the computed synchrotron age of the source and the ratio between the quiescent and the active timescales; Column 7: the reduced chi-square of the fit.
}}
\end{tabular}
}
\label{tab:nub-chi}
\end{table}

By spending 30\,percent of its lifetime in the quiescent phase, the $t_{\rm off}/t_{\rm s}$ ratio of J1615+5452 fits within the range $0.2-0.8$ reported in \citet{2007A&A...470..875P} and \citet{2011A&A...526A.148M}. Once the dormant phase of the nuclear engine kicks off, either permanently or intermittently, the source compact radio structure will (partially)
disappear due to the lack of freshly injected particles.

The diffuse radio emission of the source at low frequencies is similar to the peculiar morphology of other dying radio AGNs such as B2 0924+30
and blob1 which were also discovered (partly) because of their amorphous radio emission. Assuming that the host galaxy resides in a loose group environment (see Section\,\ref{sec:host}), then J1615+5452 also resembles to these two sources in a way that they are among the few known examples of non cluster-based dying radio galaxies
\citep{2004A&A...427...79J,2014MNRAS.440.1542D,2015MNRAS.447.2468H,2019PASA...36...16D}.

Finally, based  on the different stages of the core activity (assuming the AGN lifecycle is tightly related to the accretion of matter onto the SMBH), the population of radio galaxies can be classified into three broad categories: the active, the dying and the restarted ones \citep{1987MNRAS.227..695C,2017NatAs...1..596M}. In the case of an intermittently active radio source, the nuclear engine undergoes multiple cycles of AGN activity of the order of $\sim$\,$10^5$\,yr instead of a continuous activity with a single accretion phase that usually lasts $10^7 - 10^9$yr \citep{2004MNRAS.351..169M,2015MNRAS.451.2517S}. During the quiescent AGN phase, the nuclear activity is either significantly reduced or temporarily switched off due to a shortage of material accreting onto the SMBH. The central engine of the nuclear activity will eventually reignite once a new phase of accretion onto the SMBH starts again. Such a rejuvenation of AGN activity is deemed responsible for the presence of a flat-spectrum radio core that co-exists with remnant radio lobes from past nuclear activity \citep[e.g.][]{2009BASI...37...63S, 2017NatAs...1..596M}. With no direct evidence of a restarted core and/or new born jets for the current source, and based on the derived characteristics in Sections\,\ref{sec:radio-prop} and \ref{sec:discuss}, we potentially classify J1615+5452 as a dying radio AGN. This implies that the radio core no longer fuels fresh particle injection. The fossil radio lobes, however, remain visible (up to $\sim 10^8$\,yr depending on the radio source environment) because of the radiating low energy particles being relatively unaffected by the fast spectral evolution.
 
\section{Summary and conclusions}\label{sec:conclusion}

We have presented in this paper the discovery of a candidate remnant radio galaxy J1615+5452 in the ELAIS-N1 field. Using low frequency GMRT observations combined with
archival VLA data, we derived the source flux densities between 150 and 1400\,MHz to investigate the source
radio spectrum. Our main results are summarized below: 

\begin{enumerate}

\item The candidate remnant AGN is likely associated with an early-type elliptical galaxy at redshift of $z \sim 0.33$. The potential host galaxy appears to reside in a low-density environment.

\item With an angular extent of $\sim$\,100\,kpc, J1615+5452 has a diffuse amorphous
 radio emission with no evidence of compact core, jets and hotspots. We also did not detect radio 
 counterparts in the 1.4\,GHz VLA data with a high resolution of $\sim$\,5 arcsec.  
 
\item The spectral indices between 150 and 1400\,MHz vary between $-0.6$ and $-1.6$ which give a high spectral curvature $\Delta\alpha \sim -1$. Such spectral features indicate the predominance of nonthermal synchrotron emission with strong ongoing radiative losses. 

\item The radio spectrum is reasonably described by a CI$_{\rm off}$ model with a break frequency $\nu_{\rm b} = 384$\,MHz and an off component break at $\nu_{\rm b,\,exp} \sim 5$\,GHz. The computed  synchrotron age is $t_{\rm s}=76.0\,^{+7.4}_{-8.7}$\,Myr with $t_{\rm on} = 54.9\,^{+7.4}_{-8.5}$\,Myr and $t_{\rm off} = 21.0\,^{+0.0}_{-2.1}$\,Myr. The source is estimated to have spent 30\,percent of its total lifetime in the fading phase.
\end{enumerate}

 The morphological characteristics of J1615+5452 coupled with its spectral properties and synchrotron age
 helped us to classify the peculiar source as a dying radio galaxy. Combining such analyses is key to identify elusive remnant radio AGNs that play an important role
 toward a comprehensive understanding of AGN evolution in general. With new or upgraded low frequency radio interferometers such as LOFAR and the upgraded GMRT (uGMRT, \citealp{ugmrtpaper}), ongoing analyses of deep radio surveys are expected to significantly increase the detection of these rare sources. Furthermore, prior the completion of the Square Kilometre Array\footnote{\href{https://www.skatelescope.org/}{https://www.skatelescope.org}} (SKA), instruments like the Australian SKA Pathfinder (ASKAP, \citealp{2008ExA....22..151J}) and the South African MeerKAT telescope (\citealp{2016mks..confE...1J, 2018ApJ...856..180C}) are also expected to provide GHz frequency observations with arcsec resolution and unprecedented sensitivities to large scale structures. Such capabilities are ideal for identifying remnant radio galaxies despite the source spectral evolution.

 \section*{Acknowledgements}
 
The authors thank the referee for detailed comments and constructive suggestions. We are also grateful to the staff of the GMRT that made these observations possible. GMRT is run by the National Centre for Radio Astrophysics of the Tata Institute of Fundamental Research. This work has made use of the Cube Analysis and Rendering Tool for Astronomy \citep[CARTA,][]{comrie}. ZR acknowledges the support and funding from the South African Astronomical Observatory and the South African Radio Astronomy Observatory, which are facilities of the National Research Foundation, an agency of the Department of Science and Innovation. CHIC acknowledges the support of the Department of Atomic Energy, Government of India, under the project  12-R\&D-TFR-5.02-0700. 

 \section*{Data availability}
 The data underlying this article are available in Table\,\ref{tab:prop-sources} of the article. This publication makes use of archival GMRT datasets and VLA image available at https://naps.ncra.tifr.res.in/goa/ and https://www.cv.nrao.edu/nvss/, respectively. GMRT radio images will be shared on reasonable request to the corresponding author.

\bibliographystyle{mnras}
\bibliography{J1615}

\bsp	
\label{lastpage}
\end{document}